\documentclass[%
 reprint,
superscriptaddress,
amsmath,amssymb,
aps,
prd,
floatfix,
]{revtex4-2}

\usepackage{graphicx}
\usepackage{dcolumn}
\usepackage{bm}
\usepackage{float}
\usepackage{color}
\usepackage{soul}
\usepackage{multirow}

\DeclareUnicodeCharacter{2212}{-}
\DeclareUnicodeCharacter{02BC}{'}

\begin{document}

\preprint{APS/123-QED}


\title{Probing Forward–Backward Multiplicity Correlations and Fluctuations Using the Strongly Intensive Observable $\Sigma$ in Pb--Pb Collisions at SPS Energies with UrQMD}

\author{Ekata Nandy}
\email{ekatanandy@gmail.com}
\affiliation{Variable Energy Cyclotron Centre, 1/AF, Bidhan nagar, Kolkata 700064}

\author{Subhasis Chattopadhyay}
\email{sub.chattopadhyay@gmail.com}
\affiliation{GSI Helmholtzzentrum f\"{u}r Schwerionenforschung GmbH (GSI), Darmstadt, Germany}

\date{\today}

\begin{abstract}
The strongly intensive observable $\Sigma$, constructed from charged-pion multiplicities in separated forward (F) and backward (B) pseudorapidity intervals $\Delta\eta$, is investigated within the UrQMD transport model for Pb--Pb collisions at SPS energies. The dependence of $\Sigma$ on $\Delta\eta$ is studied as a function of collision energy, centrality, and acceptance to explore the longitudinal structure of fluctuations and correlations in particle production. For $\sqrt{s_{NN}}\gtrsim6$ GeV, $\Sigma$ exhibits a non-monotonic behavior, increasing from values close to unity at small $\Delta\eta$, reaching a maximum at intermediate separation, and subsequently decreasing towards unity at large $\Delta\eta$. A decomposition of $\Sigma$ into scaled variance and covariance terms shows that this behavior originates from different $\Delta\eta$ dependences of these two terms. With resonance decays disabled, calculation of $\Sigma$  reveals that resonance contributions dominate the short-range component of the correlations, while the persistence of $\Sigma>1$ at larger $\Delta\eta$ indicates fluctuation and correlation effects extending over broader pseudorapidity intervals. The magnitude of $\Sigma$ as a function of $\Delta\eta$, including fluctuation and covariance, increases from central to peripheral collisions,  with fluctuation dominating over covariance at large $\Delta\eta$. With collision energy increasing, the pseudorapidity interval over which $\Sigma$ remains above unity becomes significantly larger. At lower energies, where resonance dynamics completely dominate particle production, $\Sigma$ exhibits an approximately flat dependence on $\Delta\eta$. An acceptance-scaling study reveals deviations from simple binomial scaling in regions dominated by strong correlations. These results demonstrate the sensitivity of $\Sigma$ to the interplay between multiplicity fluctuations and F--B correlations in heavy-ion collisions.
\end{abstract}


\maketitle


\section{\label{sec:level1}Introduction}
Forward--backward (F--B) multiplicity correlations provide an effective tool for investigating the mechanisms of particle production in high-energy hadronic and nuclear collisions~\cite{Uhlig1978, UA5_1983, UA5_1988, E735_1995, STAR_2009, ATLAS_2012}. In such collisions, particles are produced through a variety of microscopic processes, including resonance decays, cluster formation, and the fragmentation of color strings. These mechanisms generate correlations among particles emitted in different regions of phase space. By studying how particle multiplicities in separated pseudorapidity intervals ($\Delta\eta$) are correlated, one can gain insight into the dynamics of particle production as well as the spatial and temporal structure of the underlying processes~\cite{Capella1978, Amelin1994, Braun2000, Braun2004, Brogueira2007, Armesto2007, Vechernin2007, Konchakovski2009, Lappi2010, Bzdak2012, Olszewski2013, Vechernin2004, Vechernin2011}. For example, multiple parton interactions are expected to generate long-range correlations (LRC) extending beyond $|\eta| \sim 1$ in nuclear collisions compared to hadronic interactions at the same energy~\cite{Walker2004}. Similarly, within the Color Glass Condensate (CGC) framework, correlations originating at early times in the collision can extend over large pseudorapidity intervals, whereas particles produced at later stages exhibit more localized correlations~\cite{Kovchegov:1999ep}.

A particularly useful observable for such studies is the strongly intensive quantity $\Sigma$, which characterizes the correlation between particle multiplicities measured in forward and backward pseudorapidity($\eta$) windows~\cite{PhysRevC.84.014904}. Strongly intensive observables are constructed to minimize the influence of trivial volume fluctuations arising from event-by-event variations in the system size or in the number of particle production sources. Consequently, $\Sigma$ provides a robust measure of genuine dynamical correlations between the two $\eta$-intervals.

The observable $\Sigma[N_F,N_B]$ used in this work is defined as
\begin{equation}
\Sigma[N_F,N_B] =
\frac{1}{C_\Sigma}
\begin{aligned}
\Big[
&\langle N_B \rangle \omega_{N_F}
+ \langle N_F \rangle \omega_{N_B} \\
&- 2\mathrm{Cov}(N_F,N_B)
\Big]
\end{aligned}
\label{eq:sigma_def}
\end{equation}

where
\begin{equation}
\omega_X =
\frac{\langle X^2 \rangle - \langle X \rangle^2}{\langle X \rangle}
\label{eq:scaled_variance}
\end{equation}
is called scaled variance and
\begin{equation}
C_\Sigma = \langle N_F \rangle + \langle N_B \rangle .
\label{eq:normalization}
\end{equation}

The strongly intensive nature of $\Sigma$ can be demonstrated explicitly within an independent superposition model, in which particle production arises from a fluctuating number of identical and independent sources. In this framework, the forward and backward multiplicities can be written as
\begin{equation}
N_F = \sum_{i=1}^{k} n_{F,i}, \qquad
N_B = \sum_{i=1}^{k} n_{B,i},
\label{eq:superposition}
\end{equation}
where $n_{F,i}$ and $n_{B,i}$ denote the multiplicities produced by a single source in the forward and backward windows, respectively, and $k$ is the number of sources. The corresponding mean multiplicities are
\begin{equation}
\langle N_F \rangle = \langle k \rangle \langle n_F \rangle, \qquad
\langle N_B \rangle = \langle k \rangle \langle n_B \rangle .
\label{eq:means}
\end{equation}

The scaled variances of $N_F$ and $N_B$ can be expressed as
\begin{equation}
\omega_{N_F} = \omega_{n_F} + \langle n_F \rangle \omega_k, \qquad
\omega_{N_B} = \omega_{n_B} + \langle n_B \rangle \omega_k,
\label{eq:scaled_variances}
\end{equation}
where $\omega_k = \mathrm{Var}(k)/\langle k \rangle$ denotes the scaled variance of the number of sources. The covariance between $N_F$ and $N_B$, $\mathrm{Cov}(N_F,N_B)$ is given by 
\begin{equation}
\langle N_F N_B \rangle - \langle N_F \rangle \langle N_B \rangle
=
\langle k \rangle \mathrm{Cov}(n_F,n_B)
+
\langle n_F \rangle \langle n_B \rangle \mathrm{Var}(k).
\label{eq:covariance}
\end{equation}

Substituting Eqs.~(\ref{eq:means})--(\ref{eq:covariance}) into Eq.~(\ref{eq:sigma_def}), all terms proportional to $\mathrm{Var}(k)$ cancel exactly, yielding
\begin{equation}
\Sigma[N_F,N_B]
=
\frac{
\langle n_B \rangle \omega_{n_F}
+
\langle n_F \rangle \omega_{n_B}
-
2\,\mathrm{Cov}(n_F,n_B)
}{
\langle n_F \rangle + \langle n_B \rangle
}.
\label{eq:sigma_final}
\end{equation}

Equation~(\ref{eq:sigma_final}) demonstrates explicitly that $\Sigma[N_F,N_B]$ is independent of both the average number of sources $\langle k \rangle$ and its event-by-event fluctuations. Consequently, $\Sigma$ provides a robust measure of genuine F--B correlations, eliminating trivial volume effects~\cite{Sputowska:2023wdr, Sputowska:2022gai}.

The dependence of $\Sigma$ on the $\eta$-separation between the forward and backward windows provides insight into the characteristic range of these correlations. At small separations, correlations are dominated by short-range effects (SRC) arising primarily from resonance decays and cluster formation, which produce particles correlated over limited $\eta$-intervals. At larger separations, correlations may persist due to mechanisms associated with the fragmentation of color strings formed in the early stages of the collision. Such strings can extend over wide $\eta$ range and produce particles in widely separated regions of phase space. In addition, global conservation laws, such as electric charge, baryon number, and strangeness conservation, introduce event-wide constraints that can also influence F--B correlations.

The study of these effects is particularly relevant at Super Proton Synchrotron (SPS) energies, where the interplay between different particle production mechanisms can be explored in detail. In this regime, the overall particle multiplicity is moderate compared to higher energies, enhancing the relative importance of resonance decays, cluster production, and global conservation laws. At the same time, string fragmentation is proposed to be a significant source of particle production, allowing both SR and LR correlations to be investigated within the same framework. Furthermore, SPS energies correspond to the region where key phenomena, such as the onset of deconfinement and changes in particle production dynamics, are expected to occur. Measurements of F--B multiplicity correlations in this energy domain therefore provide valuable insight into the transition between different regimes of particle production.

In this context, studying the dependence of $\Sigma$ on the $\eta$-separation ($\Delta\eta$) between forward and backward $\eta$-windows provides a sensitive probe of the relative contributions of short- and long-range correlations. The behavior of this observable can thus be used to understand the characteristic correlation lengths and to improve our understanding of the underlying dynamics governing particle production at these energies.

The organization of the paper is as follows. In section II, we give an account of UrQMD model and analysis method, followed by results \& discussion and, summary in section III and IV, respectively.

\section{Event Generator: UrQMD}
\begin{figure*}[t]
	\centering 
	\includegraphics[height=0.52\textheight, width=0.98\textwidth, angle=0]{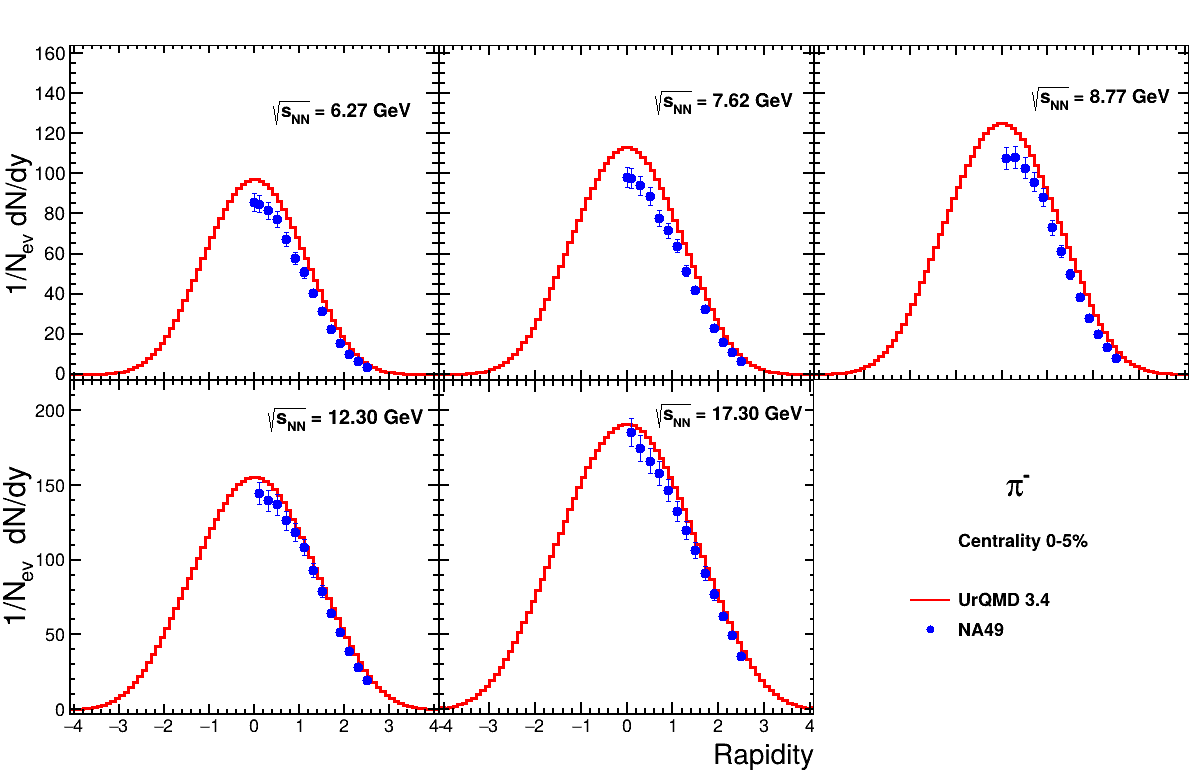}	
	\caption{(Color online) Rapidity distributions of negatively charged pions ($\pi^-$) in 0--5\% central Pb--Pb collisions at SPS energies. The distributions obtained from UrQMD 3.4 are compared with NA49 data~\cite{NA49:2002pzu, NA49:2007stj}, demonstrating that the model provides a reasonable description of $\pi^{-}$ production over the studied energy range} 
	\label{y_piMinus}%
\end{figure*}

The Ultra-relativistic Quantum Molecular Dynamics (UrQMD) model is a microscopic transport approach widely used to describe hadronic and nuclear collisions over a broad range of energies~\cite{Bass1998, Bleicher1999}. It is based on the covariant propagation of hadrons and incorporates both elastic and inelastic binary interactions, resonance excitation and decay, and string excitation and fragmentation processes. 

At low collision energies, particle production in UrQMD is dominated by the excitation and subsequent decay of baryonic and mesonic resonances. As the collision energy increases, string degrees of freedom become increasingly important, leading to particle production via string excitation and fragmentation mechanisms. The model includes a large set of hadronic states and cross-sections, allowing for a realistic description of the space--time evolution of the system. UrQMD treats the collision as a sequence of individual hadron--hadron interactions, where particles propagate along classical trajectories between successive collisions. 

As a first step to test the validity of the model, we compare the rapidity distributions of $\pi^-$ in the SPS energy range. A good description of the single-particle rapidity (or $\eta$) distributions is essential, as F--B correlation observables are constructed from event-by-event multiplicities within specified $\eta$-intervals. 

The comparison between the model calculations and available experimental data from NA49 shows that UrQMD reproduces the overall shape and magnitude of the $\pi^-$ rapidity spectra reasonably well across the considered energy range~\cite{NA49:2002pzu, NA49:2007stj}. In particular, the model captures the characteristic features of the distributions, such as the magnitude at midrapidity and the fall-off towards forward and backward rapidities. Minor deviations observed, which are within 10-20\%,  may not significantly affect the integrated multiplicities within the acceptance windows used for this F--B correlation study. 

\subsection{Analysis method}
\begin{figure}[t]
	\includegraphics[height=0.25\textheight, width=0.4\textwidth, angle=0]{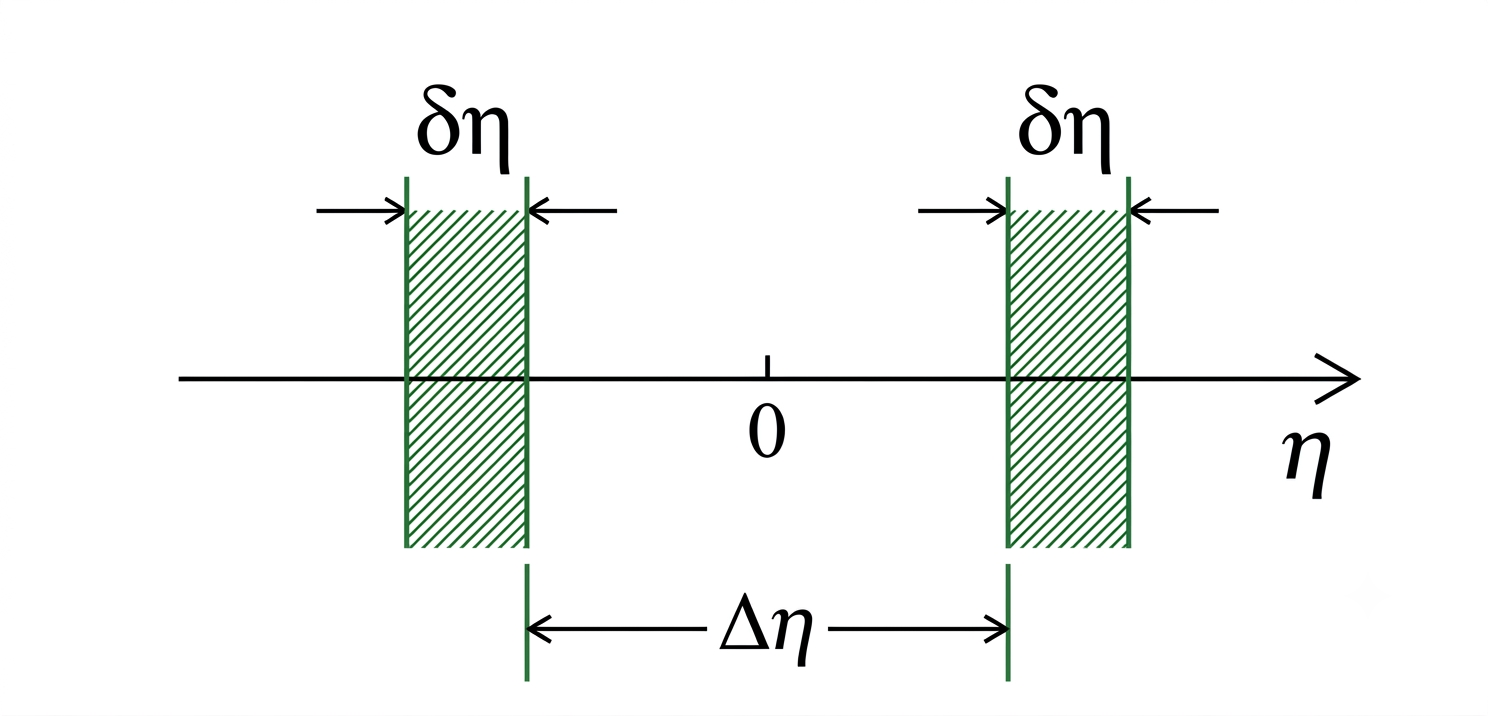}	
	\caption{(Color online) Schematic illustration of the forward (F) and backward (B) pseudorapidity windows used in the analysis, showing the window width $\delta\eta$ and the pseudorapidity separation $\Delta\eta$ between the nearest window edges.} 
	\label{illustationfig}%
\end{figure}
The F--B correlation is studied for the $0$--$5\%$ most central Pb-Pb collisions at SPS energy range 6.3 to 17.3 GeV. At each energy almost 500K events were generated in impact parameter range 0 $<$ b $<$ 3.5 fm, that corresponds to approximately $0$--$5\%$ central collision. The forward  and backward pseudorapidity windows are chosen symmetrically around $\eta = 0$, each with a fixed width of $\delta\eta = 0.3$, as shown in Fig~\ref{illustationfig}. The separation between the $F$ and $B$ windows is varied systematically by shifting their positions such that the gap between their nearest edges ranges from $0.2$ to $5.0$ in $\eta$. All charged pions with transverse momentum 0.05 $<$ p$_{T} <$ 4.0 GeV were considered for calculations. Only charged pions are considered to reduce the complexity associated baryon stopping and baryon transport phenomena particularly, at lower collision energies.

The statistical uncertainty of $\Sigma$ was estimated using the bootstrap resampling technique. For each $\Delta\eta$ interval, 100 bootstrap samples were generated by randomly selecting events from the original dataset with replacement. Each bootstrap sample contains the same number of events as the original dataset. For every bootstrap set, the observable $\Sigma$ was recalculated following the same procedure used for the original sample. The standard deviation of the bootstrap distribution of $\Sigma$ values was then used to determine the statistical uncertainty. 

\section{Results}

\begin{figure}[t]
	\centering 
	
    \includegraphics[width=0.48\textwidth, angle=0]{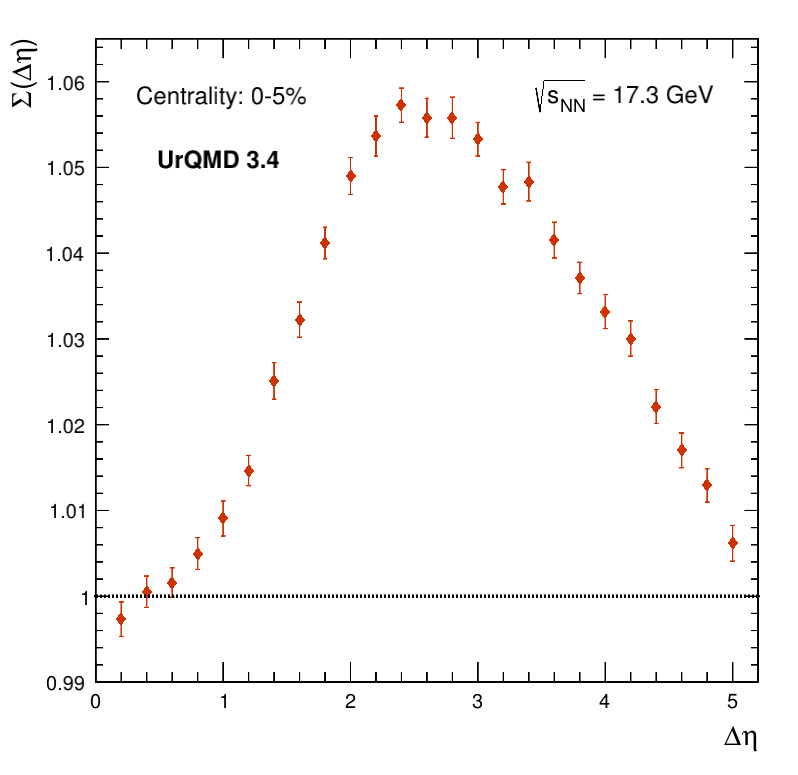}
   
	\caption{(Color online)  Strongly intensive observable $\Sigma(\Delta\eta)$ as a function of the $\Delta\eta$  separation between forward and backward windows for charged pions in 0--5\% central Pb--Pb collisions at $\sqrt{s_{\rm NN}} = 17.3$ GeV from UrQMD 3.4. The window width is fixed at $\delta\eta = 0.3$. } 
	\label{SigmaVsDeltaEta}%
\end{figure}

 For symmetric $\eta$ windows around midrapidity, the strongly intensive observable $\Sigma[N_{F},N_{B}] (\Delta\eta)$ is expressed as
 
\begin{equation}
\Sigma[N_{F},N_{B}](\Delta\eta)=\frac{\omega_F+\omega_B}{2}-\frac{\mathrm{Cov}(\Delta\eta)}{\langle F\rangle},
\end{equation}

where $\omega_F$ and $\omega_B$ denote the scaled variances of multiplicity distributions in the F and B windows, respectively, $\mathrm{Cov}(\Delta\eta)$ represents their covariance, and $\langle F\rangle=\langle B\rangle$ for symmetric configurations. By construction, $\Sigma=1$ corresponds to independent particle production limit, while deviations from unity indicate the presence of correlations and/or non-Poissonian multiplicity fluctuations. For simplicity of notation, $\Sigma[N_{F},N_{B}] (\Delta\eta)$ will be referred to as $\Sigma(\Delta\eta)$ in rest of the texts and in figures.

Figure~\ref{SigmaVsDeltaEta} presents the dependence of $\Sigma$ on the separation $\Delta\eta$ between F and B windows of fixed width $\delta\eta = 0.3$. A clear non-monotonic behavior is observed. At small separations ($\Delta\eta \approx 0$), $\Sigma$ remains close to unity. With increasing $\Delta\eta$, $\Sigma$ rises, reaches a maximum at intermediate separations, and subsequently decreases, gradually approaching unity at large $\Delta\eta$.

\begin{figure}[t]
	\centering 
	
    \includegraphics[width=0.48\textwidth, angle=0]{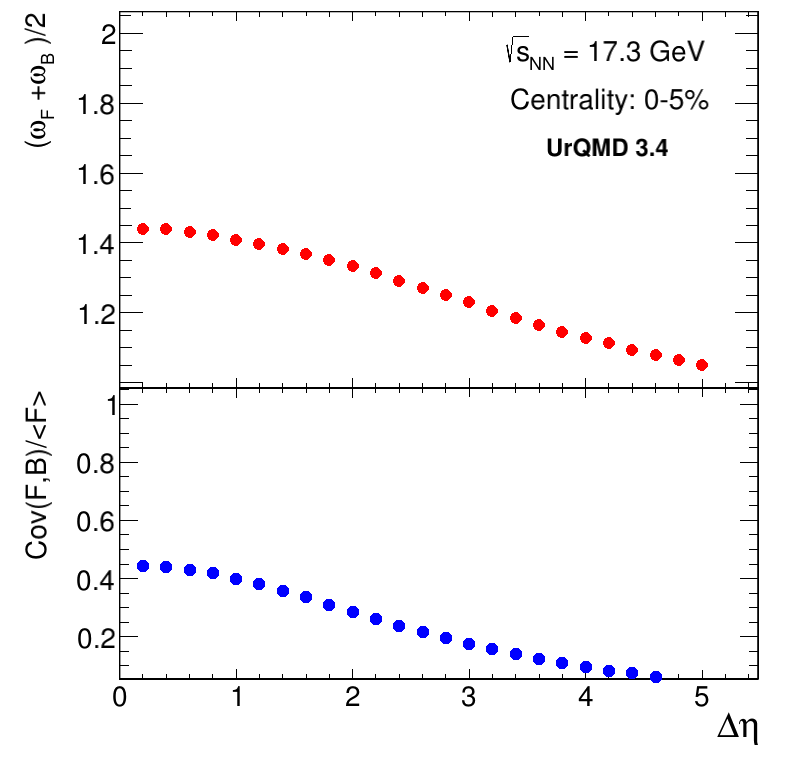}
   
	\caption{(Color online) Scaled variance $\omega_{F/B}$ and normalized F--B covariance $\mathrm{Cov}(N_F,N_B)/\langle N_F\rangle$ as a function of $\Delta\eta$ for charged pions in 0--5\% central Pb--Pb collisions at $\sqrt{s_{\rm NN}}=17.3$ GeV from UrQMD 3.4. The $\eta$-window width is fixed at $\delta\eta=0.3$. The different $\Delta\eta$ dependences of the fluctuation and covariance terms are responsible for the characteristic non-monotonic behavior observed in $\Sigma(\Delta\eta)$.}  
	\label{OmegaCovVsDeltaEta}%
\end{figure}

\begin{figure}[t]
	\centering 
	
    \includegraphics[width=0.48\textwidth, angle=0]{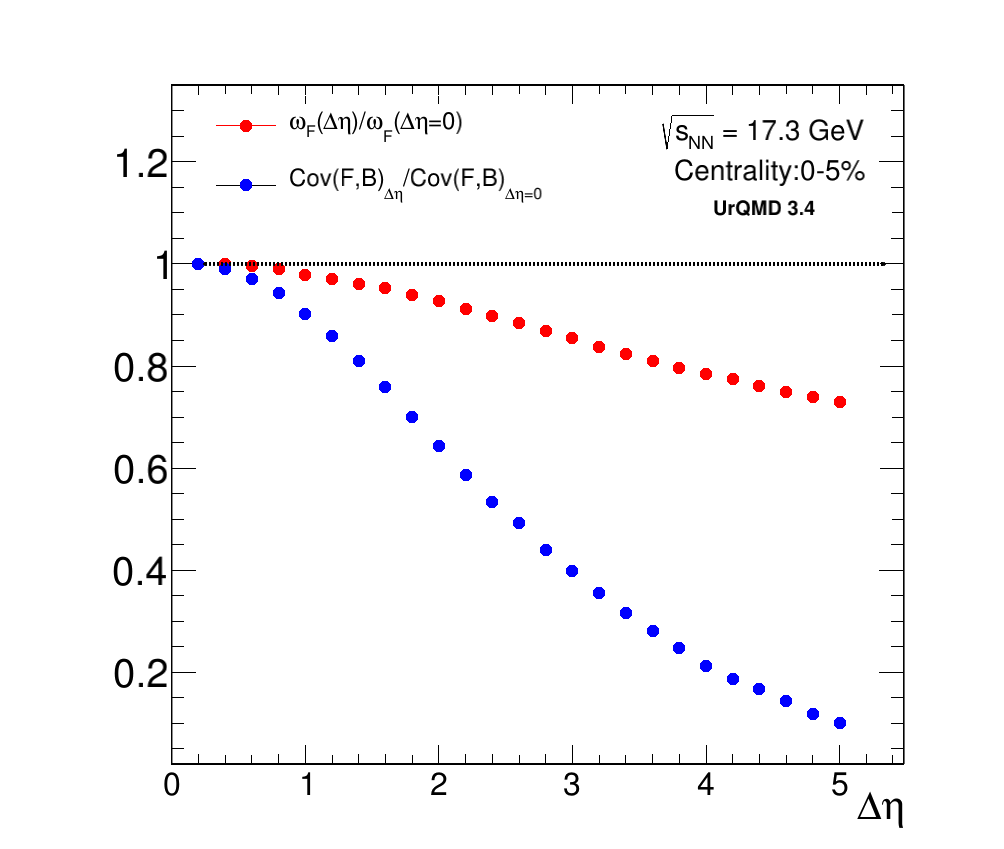}
   
	\caption{(Color online) Scaled variance $\omega_{F/B}$ and normalized forward--backward covariance $\mathrm{Cov}(N_F,N_B)/\langle N_F\rangle$ as a function of pseudorapidity separation $\Delta\eta$ for charged pions in 0--5\% central Pb--Pb collisions at $\sqrt{s_{\rm NN}}=17.3$ GeV from UrQMD 3.4. Both quantities are normalized to their respective values at $\Delta\eta=0$ in order to emphasize their relative variation with $\Delta\eta$.}
	\label{NormOmegaCovVsDeltaEta}%
\end{figure}

To understand the origin of this behavior, the individual terms in Eq.~(1) are examined in Fig.~\ref{OmegaCovVsDeltaEta}. Both the average scaled variance $(\omega_F+\omega_B) /2$ and covariance $\mathrm{Cov}(\Delta\eta)/\langle F \rangle$  decrease with increasing $\Delta\eta$. However, their normalized distributions, i.e,  scaled by their respective values at $\Delta\eta=0$, shown in Fig.~\ref{NormOmegaCovVsDeltaEta} reveal that the covariance term decreases significantly faster than the scaled variance. At small $\Delta\eta$, the covariance is large due to short-range correlations arising from resonance decays, local charge conservation, and cluster-like particle production. Since the covariance enters Eq.~(1) with a negative sign, it suppresses the contribution from multiplicity fluctuations, keeping $\Sigma(\Delta\eta)$ close to unity.

As the separation between the F and B windows increases, the probability that both decay products populating F and B windows simultaneously decreases rapidly. Thus the SR contributions to the covariance term rapidly diminishes. Consequently, the covariance term decreases more rapidly than the fluctuation term, leading to an increase of $\Sigma(\Delta\eta)$. The maximum observed at intermediate $\Delta\eta$ therefore reflects the different $\Delta\eta$ dependence of the covariance and fluctuation contributions.

At larger $\eta$-separations, the covariance becomes small and varies only weakly with $\Delta\eta$, while the scaled variance continues to decrease toward unity asymptotically. As a result, the evolution of $\Sigma(\Delta\eta)$ becomes increasingly governed by the fluctuation term, causing $\Sigma(\Delta\eta)$ to decrease and gradually approach unity. Examining the covariance term alone shows that, although the SRC contribution is strongly suppressed with increasing $\Delta\eta$, the covariance does not vanish completely even at the largest separations studied. This indicates the presence of residual correlations that persist over a broad $\eta$-range. Such residual correlations are consistent with particle-production mechanisms acting over larger longitudinal phase-space intervals in the model, for example those associated with the fragmentation of longitudinally extended color strings.
\begin{figure}[t]
	\centering 
	
    \includegraphics[width=0.48\textwidth, angle=0]{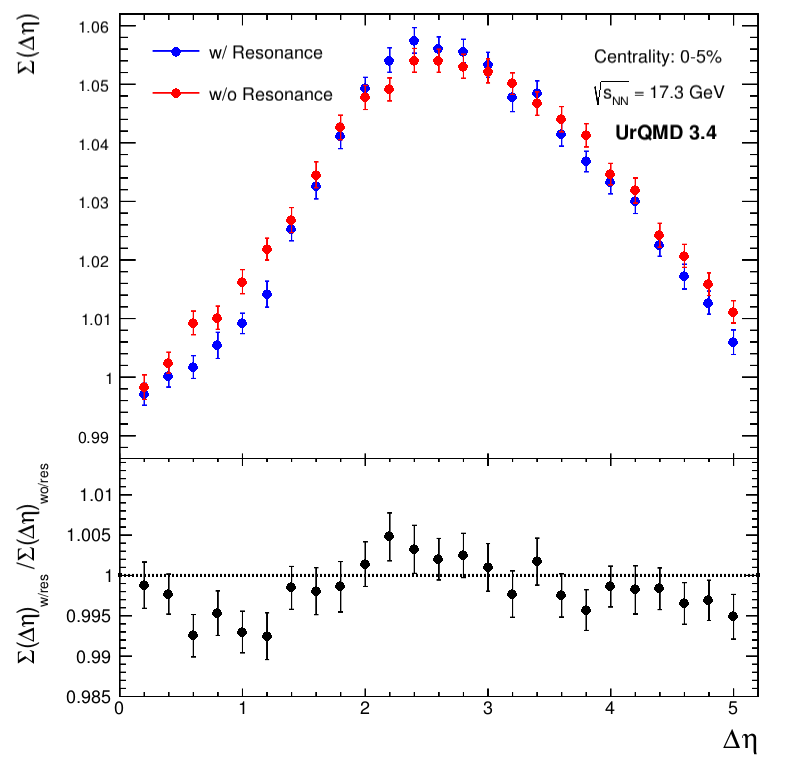}
   
	\caption{(Color online) Top panel: Dependence of the strongly intensive observable $\Sigma(F,B)$ on the $\Delta\eta$ separation between forward and backward windows for charged pions in 0--5\% central Pb--Pb collisions at $\sqrt{s_{\rm NN}}=17.3$ GeV from UrQMD 3.4. Results obtained with the default resonance decays enabled are compared to calculations with the $\rho$ and $\phi$ resonance decays switched off. The pseudorapidity window width is fixed at $\delta\eta=0.3$. Bottom panel: Ratio of $\Sigma(\Delta\eta)$ obtained with resonance decays enabled to that obtained with $\rho$ and $\phi$  resonance decays disabled.}
	\label{SigmaVsDeltaEtaResOnOff}%
\end{figure}

To further investigate the role of resonance decays, the analysis is repeated using UrQMD events with the $\rho$ and $\phi$ resonance decays switched off. The comparison is shown in Fig.~\ref{SigmaVsDeltaEtaResOnOff}. A modest but systematic increase of $\Sigma(\Delta\eta)$ is observed at small $\Delta\eta$ when resonance decays are disabled. This behavior is consistent with the expectation that resonance decays contribute predominantly to SRC. In particular, decays such as $\rho \rightarrow \pi^+\pi^-$ generate correlated pion pairs that enhance the $\mathrm{Cov}(\Delta\eta)$ at small $\Delta\eta$. Since in definition of $\Sigma$ the covariance term has a negative sign, the reduction of the covariance term in the resonance-suppressed sample leads to larger values of $\Sigma$.

The difference between $\Sigma(\Delta\eta)$ calculated with and without resonance decays rapidly diminishes with increasing $\Delta\eta$ and becomes negligible at intermediate and large $\eta$-separations, as shown by the ratio plot in the bottom panel of Fig.~\ref{SigmaVsDeltaEtaResOnOff}. This observation supports the interpretation that resonance decays contribute predominantly to the short-range component of the F--B correlations. Furthermore, the persistence of non-zero values of $\mathrm{Cov}(\Delta\eta)$ at large $\Delta\eta$ cannot be attributed solely to resonance feed-down. These results therefore indicate the presence of additional correlation mechanisms acting over a broader $\eta$-range within the model, consistent with long-range correlation effects.

\begin{figure}[h]
	\centering 
	
    \includegraphics[width=0.48\textwidth, angle=0]{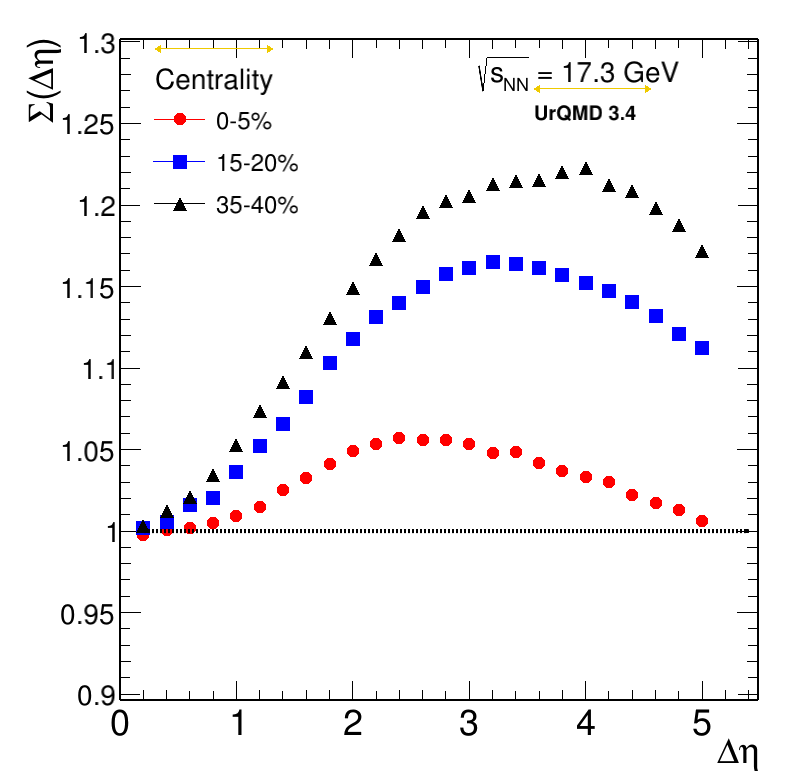}
   
	\caption{(Color online) Dependence of $\Sigma(\Delta\eta)$ on the pseudorapidity separation between the forward and backward windows for charged pions in Pb--Pb collisions at $\sqrt{s_{\rm NN}}=17.3$ GeV, calculated using UrQMD 3.4. Results are shown for three centrality intervals: 0--5\%, 15--20\%, and 35--40\%. The width of the forward and backward pseudorapidity windows is fixed at $\delta\eta=0.3$. A systematic increase in the magnitude of $\Sigma$ and a shift of the maximum towards larger $\Delta\eta$ are observed from central to peripheral collisions.} 
	\label{SigmaVsDeltaEtaCentrality}%
\end{figure}

Figure~\ref{SigmaVsDeltaEtaCentrality} shows the
centrality dependence of $\Sigma(\Delta\eta)$. 
The magnitude of $\Sigma(\Delta\eta)$ increases progressively from central to peripheral  collisions over the entire $\Delta\eta$ range. In addition, the position where $\Sigma(\Delta\eta)$ achieves the maximum shifts towards larger $\Delta\eta$ values for more peripheral event classes.

To understand the origin of this behavior, the individual contributions in Eq.~(1), i.e, $(\omega_{F} + \omega_{B})/2$ and $\mathrm{Cov}(\Delta\eta)/\langle F \rangle$  were investigated as a function of centrality. Both the average scaled variance between F and B interval and the covariance term were found to increase from central to peripheral collisions, indicating an enhancement of multiplicity fluctuations as well as F--B correlations in peripheral events. Since the microscopic particle-production mechanisms implemented in UrQMD are common to all collision centralities, the observed centrality dependence does not imply the emergence of any new mechanisms in peripheral collisions. Instead, it reflects a change in the relative strengths of the fluctuation and correlation contributions in $\Sigma$. Although both the scaled variance and the covariance increase towards peripheral collisions, the stronger increase of the fluctuation term leads to the observed enhancement of $\Sigma(\Delta\eta)$.

\begin{figure}[h]
	\centering 
	
    \includegraphics[width=0.48\textwidth, angle=0]{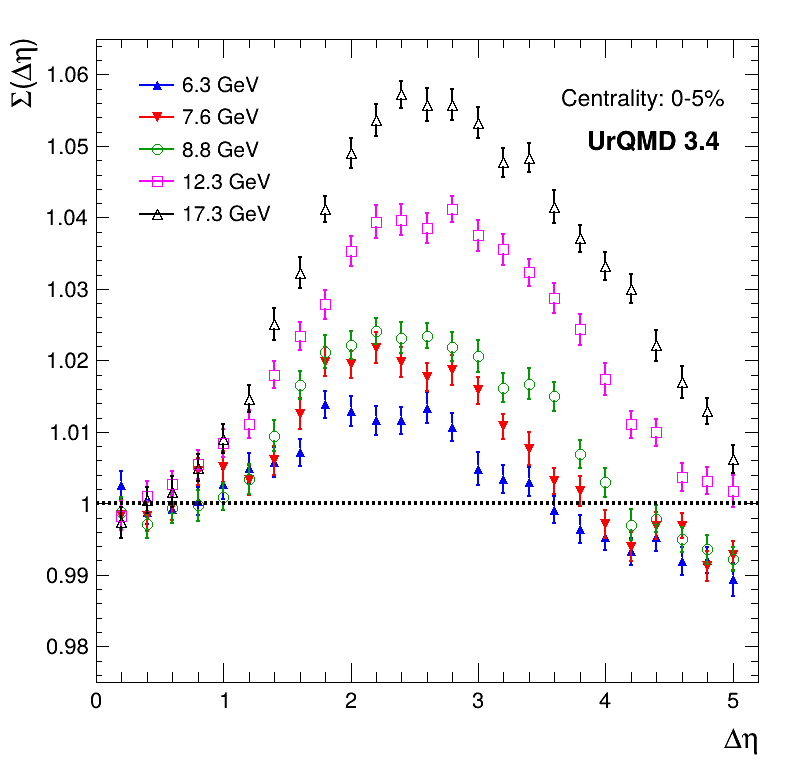}
   
	\caption{(Color online) Dependence of the strongly intensive observable $\Sigma(\Delta\eta)$ on the pseudorapidity separation between the forward and backward windows for charged pions in 0--5\% central Pb--Pb collisions at SPS energies, $\sqrt{s_{\rm NN}}=6.3$--17.3 GeV, calculated using UrQMD 3.4. The width of the forward and backward pseudorapidity windows is fixed at $\delta\eta=0.3$. With increasing collision energy, both the maximum value of $\Sigma$ and the pseudorapidity interval over which $\Sigma$ remains above unity increase systematically.
} 
	\label{SigmaVsDeltaEtaEnergy}%
\end{figure}

Next the energy dependence of $\Sigma(\Delta\eta)$ is investigated over the range $\sqrt{s_{NN}}=6.3$--17.3 GeV, corresponding to SPS beam energies from 20A to 158A GeV. A clear evolution with collision energy is observed in Fig.~\ref{SigmaVsDeltaEtaEnergy}, both in the magnitude of the maximum and in the $\Delta\eta$ at which $\Sigma$ approaches unity. At lower energies (e.g., $\sqrt{s_{NN}}=$6.3) GeV), $\Sigma$ returns to values close to unity at $\Delta\eta\approx$3, whereas at $\sqrt{s_{NN}}=$17.3 GeV this occurs only at substantially larger separations, $\Delta\eta\approx$5.

\begin{figure}[h]
	\centering 
	
    \includegraphics[width=0.48\textwidth, angle=0]{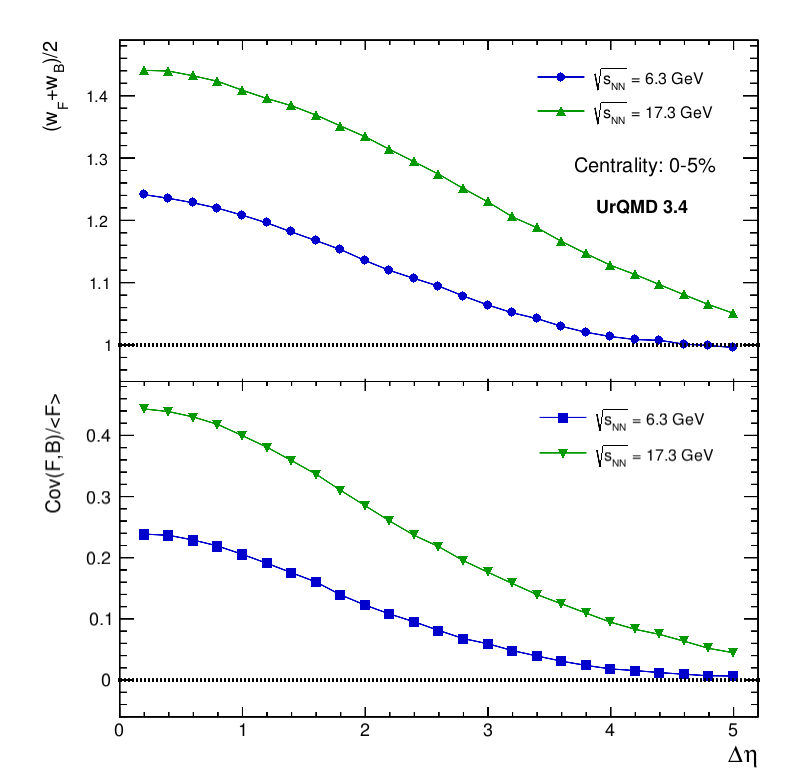}
   
	\caption{(Color online) Comparison of the scaled variance $\omega_{F}$ and the normalized forward--backward covariance $\mathrm{Cov}(N_F,N_B)/\langle N_F\rangle$ as a function of pseudorapidity separation $\Delta\eta$ for charged pions in 0--5\% central Pb--Pb collisions at $\sqrt{s_{\rm NN}}=6.3$ and 17.3 GeV, calculated using UrQMD 3.4. The width of the forward and backward pseudorapidity windows is fixed at $\delta\eta=0.3$. While the forward--backward covariance decreases rapidly with increasing $\Delta\eta$ at both energies, the scaled variance $for \sqrt{s_{\rm NN}}=17.3$ remains significantly above unity over a broader pseudorapidity interval than $\sqrt{s_{\rm NN}}=6.3$ GeV.
} 
	\label{w_and_cov}%
\end{figure}
To understand the origin of this behavior, the individual contributions from scaled variance and the covariance term in Eq.~(1) are examined in Fig.~\ref{w_and_cov} for lowest and highest energy i.e, $\sqrt{s_{NN}}=$ 6.3 and 17.3 GeV. At all energies, the F--B covariance decreases with increasing $\Delta\eta$ separation and remains small at large $\Delta\eta$, although a finite positive covariance persists throughout the studied energy range. In contrast, the scaled variances $\omega_{F/B}$ at the lowest collision energies, approach unity at large $\Delta\eta$, indicating multiplicity fluctuations close to the independent particle-production limit. At higher energies, however, $\omega_{F/B}$ remain above unity even at the largest $\Delta\eta$ considered.

These observations indicate that the extension of the $\Sigma>1$ region with increasing collision energy reflects the combined effect of a finite positive covariance and enhanced multiplicity fluctuations that persist over larger longitudinal phase-space intervals. Within the framework of UrQMD, the microscopic particle-production mechanisms remain the same throughout the studied SPS energy range, however, their relative contribution and longitudinal extent evolve with collision energy. With $\omega_{F/B}$ remaining above unity at high energies suggests that fluctuations associated with particle production remain significant over broader $\eta$-intervals, while the finite covariance indicates the presence of residual correlations consistent with long range mechanisms like the one from the fragmentation longitudinally stretched strings. Consequently, the observed broadening of the $\Sigma>1$ region reflects an increasing longitudinal extent of the underlying fluctuation and correlation structure associated with the particle-production mechanism.

\begin{figure}[h]
	\centering 
	
    \includegraphics[width=0.48\textwidth, angle=0]{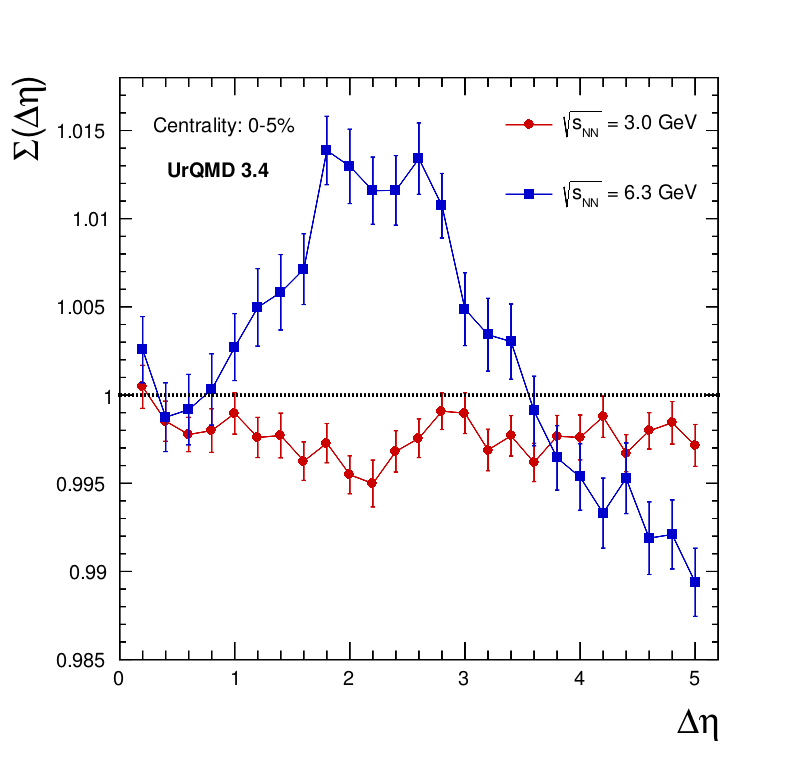}
   
	\caption{(Color online) Strongly intensive observable $\Sigma(\Delta\eta)$ as a function of pseudorapidity separation for charged pions in 0--5\% central Pb--Pb collisions at $\sqrt{s_{\rm NN}}=3.0$ and 6.3 GeV, calculated using UrQMD 3.4. The forward and backward pseudorapidity windows have a fixed width of $\delta\eta=0.3$. While $\Sigma(\Delta\eta)$ exhibits an approximately flat dependence and remains close to unity at $\sqrt{s_{\rm NN}}=3.0$ GeV, a pronounced non-monotonic behavior develops at $\sqrt{s_{\rm NN}}=6.3$ GeV, reflecting the onset of additional fluctuation and correlation mechanisms beyond resonance-dominated particle production.
}
	\label{sigma_3GeV_6GeV}%
\end{figure}

Previously, it was argued that the strongly intensive observable $\Sigma(F,B)$ is sensitive to the underlying particle-production dynamics. To further explore this aspect, $\Sigma(\Delta\eta)$ is compared for $\sqrt{s}=$3 GeV and $\sqrt{s}=$6.3 GeV. Within the UrQMD framework, particle production at $\sqrt{s}\lesssim$5 GeV is dominated by resonance excitation and decay processes, while contributions from string excitation and fragmentation become increasingly important at higher energies. Consequently, correlation mechanisms extending over large rapidity intervals are expected to be significantly reduced at $\sqrt{s}=$3 GeV.

Figure~\ref{sigma_3GeV_6GeV} presents the corresponding comparison. At $\sqrt{s}=$3 GeV, $\Sigma$ exhibits an approximately flat dependence on $\Delta\eta$, with values remaining slightly below unity over the entire $\Delta\eta$ range. This behavior is consistent with the results discussed previously, where the scaled variances approach values close to unity while a small positive F--B covariance persists, leading to $\Sigma<$1. 
In contrast, at $\sqrt{s}=$6.3) GeV, $\Sigma$ develops a clear non-monotonic dependence on $\Delta\eta$, characterized by an increase from values close to unity, the appearance of a maximum at intermediate separations, and a gradual decrease at larger $\Delta\eta$. As discussed earlier, such behavior results from the interplay between the fluctuation and covariance. The emergence of this structure indicates the onset of additional particle-production mechanisms that generate fluctuations and correlations over broader longitudinal phase-space intervals.

The comparison between $\sqrt{s}=$3 GeV and $\sqrt{s}=$6.3 GeV therefore demonstrates that $\Sigma$ is sensitive to changes in the dominant particle-production dynamics. While resonance-dominated particle production leads to an approximately flat behavior of $\Sigma(\Delta\eta)$, the appearance of a non-monotonic $\Delta\eta$ dependence at higher energies reflects the growing importance of mechanisms capable of producing correlations and fluctuations over extended rapidity intervals. This highlights the utility of $\Sigma$ as a probe of the underlying particle-production processes in heavy-ion collisions.

\begin{figure}[h]
	\centering 
	
    \includegraphics[width=0.48\textwidth, angle=0]{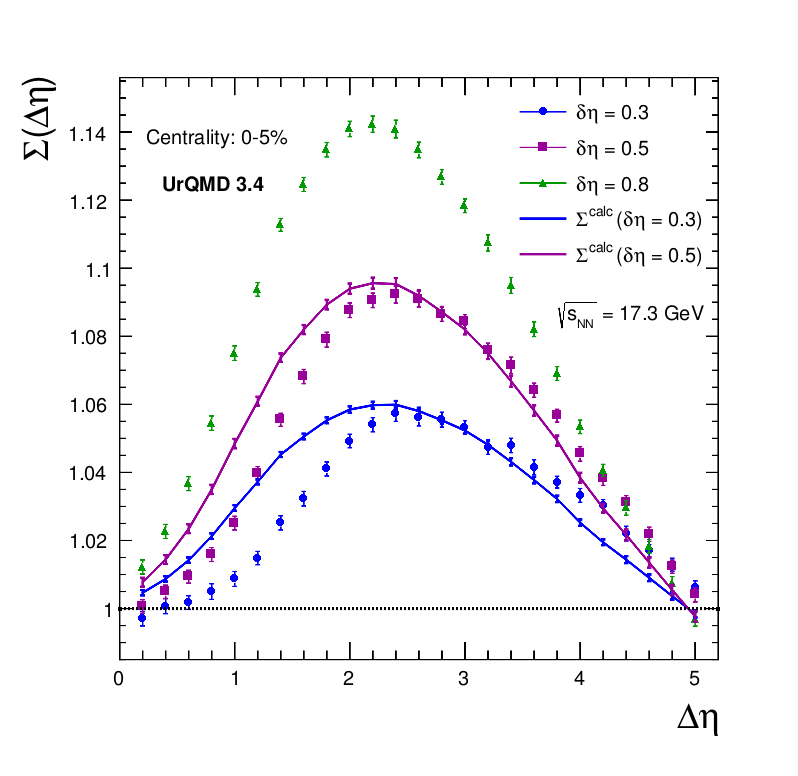}
   
	\caption{(Color online) Dependence of the strongly intensive observable $\Sigma(\Delta\eta)$ on pseudorapidity separation for charged pions in 0--5\% central Pb--Pb collisions at $\sqrt{s_{\rm NN}}=17.3$ GeV calculated using UrQMD 3.4. The directly calculated results for pseudorapidity window widths $\delta\eta=0.3$, 0.5, and 0.8 are shown together with the values for $\delta\eta=0.3$ and 0.5 obtained from the finite-acceptance scaling relation of Eq.~(23), using the $\delta\eta=0.8$ measurement as the reference. The figure illustrates the extent to which the binomial acceptance-scaling prescription reproduces the measured $\Sigma(\Delta\eta)$ over the studied pseudorapidity range.
} 
	\label{sigma_calc_vs_scaling}%
\end{figure}
It is well known that correlation and fluctuation measurements are sensitive to acceptance. When the phase-space coverage is limited, only a fraction of the produced particles is sampled, leading to a partial capture of the underlying correlation or fluctuation compared to measurements performed over a wider acceptance. It is therefore important to quantify how the strongly intensive observable $\Sigma$ is affected by acceptance.

To investigate this, $\Sigma$ is evaluated for different $\delta\eta$ windows. In particular, results obtained for smaller windows, $\delta\eta = 0.3$ and $0.5$, are compared to those from a larger reference window $\delta\eta = 0.8$. Since the true value of 
$\Sigma$ corresponding to full phase-space coverage is not practically accessible, the result obtained in the large acceptance is used as an effective reference. The scaling relation is then tested by comparing measurements at smaller acceptances to the expectation derived from this reference using the scaling relation.

To derive the scaling relation, the multiplicities in the forward and backward windows, $F_{\mathrm{acc}}$ and $B_{\mathrm{acc}}$, are treated as binomial samples of the true multiplicities $F$ and $B$:

\begin{equation}
F_{\mathrm{acc}} \sim \mathrm{Binomial}(F,\alpha), \qquad
B_{\mathrm{acc}} \sim \mathrm{Binomial}(B,\alpha),
\end{equation}
where $\alpha$ denotes the probability that produced particle is detected within the acceptance.

The mean multiplicities in the accepted region are then
\begin{equation}
\langle F_{\mathrm{acc}} \rangle = \alpha \langle F \rangle, \qquad
\langle B_{\mathrm{acc}} \rangle = \alpha \langle B \rangle.
\end{equation},

For binomial sampling, the variance of the accepted multiplicity becomes

\begin{equation}
\mathrm{Var}(F_{\mathrm{acc}}) =
\alpha^2 \mathrm{Var}(F) + \alpha(1-\alpha)\langle F \rangle ,
\end{equation}

\begin{equation}
\mathrm{Var}(B_{\mathrm{acc}}) =
\alpha^2 \mathrm{Var}(B) + \alpha(1-\alpha)\langle B \rangle .
\end{equation}


The scaled variance in the accepted region is therefore

\begin{equation}
\omega_{F,\mathrm{acc}} =
\frac{\mathrm{Var}(F_{\mathrm{acc}})}{\langle F_{\mathrm{acc}}\rangle}.
\end{equation}

Substituting the expressions above,

\begin{align}
\omega_{F,\mathrm{acc}} &=
\frac{\alpha^2 \mathrm{Var}(F) + \alpha(1-\alpha)\langle F\rangle}
{\alpha \langle F\rangle} \\
&= \alpha \frac{\mathrm{Var}(F)}{\langle F\rangle} + (1-\alpha) \\
&= \alpha \omega_F + (1-\alpha).
\end{align}

Similarly,

\begin{equation}
\omega_{B,\mathrm{acc}} = \alpha \omega_B + (1-\alpha).
\end{equation}

If the true covariance between the forward and backward multiplicities is $\mathrm{Cov}(F,B)$, then under binomial acceptance

\begin{equation}
\mathrm{Cov}(F_{\mathrm{acc}},B_{\mathrm{acc}}) = \alpha^2 \, \mathrm{Cov}(F,B).
\end{equation}

The strongly intensive observable measured within the finite acceptance and for the symmetric windows is then given by

\begin{equation}
\Sigma_{\mathrm{acc}} =
\frac{\omega_{F,\mathrm{acc}} + \omega_{B,\mathrm{acc}}}{2}
-
\frac{\mathrm{Cov}(F_{\mathrm{acc}},B_{\mathrm{acc}})}
{\langle F_{\mathrm{acc}}\rangle}.
\end{equation}

Substituting the expressions derived above,

\begin{align}
\Sigma_{\mathrm{acc}}
&=
\frac{\alpha\omega_F + (1-\alpha) + \alpha\omega_B + (1-\alpha)}{2}
-
\frac{\alpha^2 \mathrm{Cov}(F,B)}{\alpha \langle F\rangle}.
\end{align}

After simplification one obtains

\begin{equation}
\Sigma_{\mathrm{acc}} =
\alpha
\left[
\frac{\omega_F + \omega_B}{2}
-
\frac{\mathrm{Cov}(F,B)}{\langle F\rangle}
\right]
+
(1-\alpha).
\end{equation}

Recognizing the term in brackets as the true strongly intensive observable $\Sigma_{\mathrm{true}}$, we obtain

\begin{equation}
\boxed{
\Sigma_{\mathrm{acc}} = \alpha \, \Sigma_{\mathrm{true}} + (1-\alpha)
}
\end{equation}

This result shows that finite detector acceptance suppresses deviations of $\Sigma$ from unity. In the limit of very small acceptance ($\alpha \rightarrow 0$), the observable approaches the Poisson baseline $\Sigma = 1$, while for full acceptance ($\alpha = 1$) the true value of $\Sigma$ is recovered.

Figure~\ref{sigma_calc_vs_scaling} compares the directly calculated values of $\Sigma$ with those obtained from the acceptance-scaling relation. The validity of the scaling relation is found to be limited to a restricted range of $\Delta\eta$ separations. In particular, the agreement between the measured and scaled values is best in the region where $\Sigma(\Delta\eta)$ reaches its maximum.

At small $\Delta\eta$, the F and B windows are strongly influenced by SRCs arising from resonance decays and other local particle-production processes. Such correlations violate the assumption of independent binomial sampling underlying the derivation of the scaling relation and therefore lead to deviations from the expected scaling behavior.

At large $\Delta\eta$, SRCs are strongly suppressed, however, finite F--B correlations and non-Poissonian multiplicity fluctuations remain present, as demonstrated by the persistence of non-zero covariance and values of $\omega_{F/B}$ above unity. These residual fluctuation and correlation contributions do not necessarily scale linearly with acceptance and can therefore also produce deviations from the scaling expectation.

The approximate agreement observed at intermediate $\Delta\eta$ should therefore be interpreted with caution. It does not imply the absence of correlations or a regime of independent particle production. Rather, it indicates that, within this $\Delta\eta$ interval, the combined effect of fluctuations and correlations is approximately consistent with the assumptions underlying the acceptance-scaling relation. The observed agreement may therefore arise from a balance between different correlation contributions rather than from their complete disappearance.

Taken together, the $\Delta\eta$ dependence of $\Sigma$ reflects the interplay between fluctuation and correlation effects acting over different pseudorapidity scales. The increase of $\Sigma$ from values close to unity towards a maximum is associated with the progressive suppression of short-range correlation contributions, while the persistence of $\Sigma>$1 at larger separations reflects the continued presence of non-Poissonian fluctuations and residual forward--backward correlations extending over broader pseudorapidity intervals. The eventual approach of $\Sigma$ towards unity indicates a gradual reduction of these contributions as the separation between the F and B windows increases.

\section{Summary}

The strongly intensive observable $\Sigma(F,B)$ has been investigated within the UrQMD model for Pb--Pb collisions at SPS energies using charged-pions multiplicities in separated F and B $\eta$- windows. The study focused on the dependence of $\Sigma$ on $\Delta\eta$ separation at various collision energies, centralities, resonance contributions, and acceptance effects in order to understand the longitudinal structure of fluctuations and correlations in particle production.

For collision energies above $\sqrt{s_{NN}}\approx$6 GeV, $\Sigma(\Delta\eta)$ exhibits a characteristic non-monotonic dependence on the separation between the F and B windows. The observable increases from values close to unity at small $\Delta\eta$, reaches a maximum at intermediate separations, and subsequently decreases towards unity at large $\Delta\eta$. A decomposition of $\Sigma$ into its constituent scaled variance and covariance terms demonstrates that this behavior originates from the interplay between multiplicity fluctuations and F--B correlations. In particular, the covariance decreases more rapidly with increasing $\Delta\eta$ separation than the scaled variance, leading to the observed maximum in $\Sigma(\Delta\eta)$.

Dedicated calculations performed with selected resonance decays switched off reveal that resonance contributions predominantly affect the SR component of the F--B correlations. The differences between calculations with and without resonance decays are prominent in small $\Delta\eta$ separations and become negligible at intermediate and large $\Delta\eta$. This observation indicates that resonance feed-down is not responsible for the residual covariance and values of $\Sigma$ above unity observed at larger separations.

The energy dependence of $\Sigma$ shows a systematic broadening of the $\eta$ interval over which the observable remains above unity. A detailed examination of the scaled variance and covariance terms reveals that this behavior cannot be attributed solely to enhanced F--B correlations. Instead, it reflects the combined effect of finite positive covariance and multiplicity fluctuations that persist over increasingly larger longitudinal phase-space intervals with increasing collision energy. At the lowest energies studied ($<$ 5 GeV), where particle production in UrQMD is dominated by resonance excitation and decay, $\Sigma$ exhibits an approximately flat dependence on $\Delta\eta$ and remains close to or slightly below unity. The comparison between $\sqrt{s_{NN}}=$3 GeV and higher SPS energies demonstrates the sensitivity of $\Sigma$ to changes in the dominant particle-production dynamics.

A pronounced centrality dependence of $\Sigma$ is observed. The magnitude of the observable increases from central to peripheral collisions, while the position of the maximum shifts towards larger $\Delta\eta$. Both the scaled variance and covariance contributions increase towards peripheral collisions, however, the stronger increase of the fluctuation term results in the observed enhancement of $\Sigma$. This behavior indicates a change in the relative balance between fluctuation and correlation contributions with collision centrality.

The acceptance-scaling study demonstrates that the simple binomial acceptance relation reproduces the measured behavior only within a limited range of pseudorapidity separations. Deviations observed at small and large $\Delta\eta$ indicate that the assumptions underlying independent particle sampling are violated by the presence of dynamical correlations and non-Poissonian multiplicity fluctuations. These results emphasize the importance of considering acceptance effects when interpreting fluctuation observables.

Overall, the present study establishes $\Sigma(F,B)$ as a sensitive probe of the longitudinal structure of particle production. The observable provides complementary information on both multiplicity fluctuations and F--B correlations and enables the separation of short-range resonance-driven effects from broader fluctuation and correlation structures. The observed dependences on collision energy and centrality suggest that the longitudinal extent and strength of the underlying fluctuation and correlation pattern evolve significantly across the SPS energy range, providing valuable insight into the dynamics of particle production in relativistic heavy-ion collisions.

\nocite{*}

\bibliography{apssamp} 

\end{document}